\newif\iffigs\figsfalse
\figstrue

\documentstyle[psfig,11pt,a4]{article}
\iffigs
\input epsf
\else
\fi 

\begin{document}



\newcounter{subequation}[equation]

\makeatletter
\expandafter\let\expandafter\reset@font\csname reset@font\endcsname
\newenvironment{subeqnarray}
{\def\@eqnnum\stepcounter##1{\stepcounter{subequation}{\reset@font\rm
  (\theequation\alph{subequation})}}\eqnarray}%
{\endeqnarray\stepcounter{equation}}
\makeatother


\newcommand{\ga}{\alpha}
\newcommand{\gb}{\beta}
\newcommand{\gc}{\gamma}
\newcommand{\gcp}{\gamma^\prime}
\newcommand{\gd}{\delta}
\newcommand{\gep}{\epsilon}
\newcommand{\gl}{\lambda}
\newcommand{\gL}{\Lambda}
\newcommand{\gk}{\kappa}
\newcommand{\go}{\omega}
\newcommand{\gp}{\phi}
\newcommand{\gs}{\sigma}
\newcommand{\gt}{\theta}
\newcommand{\gC}{\Gamma}
\newcommand{\gD}{\Delta}
\newcommand{\gO}{\Omega}
\newcommand{\gT}{\Theta}
\newcommand{\gvp}{\varphi}


\newcommand{\be}{\begin{equation}}
\newcommand{\ee}{\end{equation}}
\newcommand{\ba}{\begin{array}}
\newcommand{\ea}{\end{array}}
\newcommand{\bea}{\begin{eqnarray}}
\newcommand{\eea}{\end{eqnarray}}
\newcommand{\bes}{\begin{eqnarray*}}
\newcommand{\ees}{\end{eqnarray*}}
\newcommand{\bsea}{\begin{subeqnarray}}
\newcommand{\esea}{\end{subeqnarray}}
\newcommand{\lra}{\longrightarrow}
\newcommand{\lms}{\longmapsto}
\newcommand{\ra}{\rightarrow}
\newcommand{\pa}{\partial}


\newcommand{\N}{\mbox{I\hspace{-.4ex}N}}
\newcommand{\C}{\mbox{$\,${\sf I}\hspace{-1.2ex}{\bf C}}}
\newcommand{\Cs}{\mbox{$\,${\sf I}\hspace{-1.2ex}C}}
\newcommand{\Z}{\mbox{{\sf Z}\hspace{-1ex}{\sf Z}}}
\newcommand{\R}{\mbox{\rm I\hspace{-.4ex}R}}
\newcommand{\1}{\mbox{1\hspace{-.6ex}1}}


\makeatletter
\newcommand{\Tr}{\mathop{\operator@font Tr}\nolimits}
\newcommand{\F}{\mathop{\operator@font F}\nolimits}
\newcommand{\G}{\mathop{\operator@font G}\nolimits}
\newcommand{\me}{\mathop{\operator@font e}\nolimits}
\newcommand{\th}{\mathop{\operator@font th}\nolimits}
\newcommand{\ch}{\mathop{\operator@font ch}\nolimits}
\newcommand{\sh}{\mathop{\operator@font sh}\nolimits}
\newcommand{\co}{\mathop{\operator@font c}\nolimits}
\newcommand{\si}{\mathop{\operator@font s}\nolimits}
\newcommand{\Ai}{\mathop{\operator@font Ai}\nolimits}
\newcommand{\Bi}{\mathop{\operator@font Bi}\nolimits}
\makeatother

\newcommand{\la}{\wi{\Tr J^2}}
\newcommand{\DI}[1]{\mbox{$\displaystyle{#1}$}}
\newcommand{\wi}[1]{\widehat{#1}}



\thispagestyle{empty}

\hbox to \hsize{%
\vtop{\hbox{      }\hbox{     }} \hfill
\vtop{\hbox{DAMTP-R-98-29}}}

\vspace*{1cm}

\bigskip\bigskip\begin{center}
{\bf \Large{Classical and Quantum Analysis of Repulsive Singularities \\
        in Four Dimensional Extended Supergravity}}
\end{center}  \vskip 1.0truecm
\centerline{{\bf I. Gaida, H. R. Hollmann\footnote{e-mail: 
    H.Hollmann@damtp.cam.ac.uk} and J. M. Stewart\footnote{e-mail: 
    J.M.Stewart@damtp.cam.ac.uk}}}
\centerline{Department of Applied Mathematics and Theoretical Physics,}
\centerline{Silver Street, Cambridge CB3 9EW, England}
\vskip 2cm
\bigskip \nopagebreak 
\begin{abstract}
\noindent
Non--minimal repulsive singularities (``repulsons'') in extended
supergravity theories are investigated. The short distance 
antigravity properties of the repulsons are tested at 
the classical and the quantum level by a scalar test--particle.
Using a partial wave expansion it is shown that the particle gets  
totally reflected at the origin. A high frequency incoming particle 
undergoes a phase shift of $\frac{\pi}{2}$. 
However, the phase shift for a low--frequency 
particle depends upon the physical data of the repulson.
The curvature singularity at a finite distance $r_h$ turns out 
to be transparent for the scalar test--particle and the coordinate 
singularity at the origin serves as a repulsive barrier at which 
particles bounce off. 

\vspace*{0.1cm}

\vspace*{0.1cm}

\end{abstract}

\newpage\setcounter{page}1

\section{Introduction}

In the last years there has been a lot of progress
in understanding black hole physics in supergravity and
string theory in $N>1$ supersymmetric vacua
(for recent reviews see \cite{Lus98}). A lot of these black hole
solutions can be interpreted as certain (non--singular)
$p$--brane solutions, too \cite{GibHorTow95}.
This opens the possiblity to obtain a
microscopic understanding of the macroscopic Bekenstein--Hawking
entropy \cite{Bek73}. Another interesting point in this
context is the appearance of massless black holes at particular
points in moduli space giving rise to
gauge symmetry enhancement or supersymmetry enhancement
\cite{Str95,Gai982}. However, in
\cite{Str95,KalLin95,FerKalStr95,BehLusSab98,Gai981} it also has been shown
that not only
attractive singularities (black holes) are stable BPS solutions
of these supersymmetric vacua, but that repulsive naked
singularities (``repulsons'') appear too.
Thus, repulsons are non--perturbative manifestations of
antigravity effects in supersymmetric vacua. 
With antigravity we mean in this context the property of these solutions 
to reflect particles with a given mass or angular momentum (\cite{KalLin95}). 
It has known for a long time that antigravity effects occur at the 
perturbative level in extended supergravity \cite{Sch79}.

It has been pointed out
in \cite{Gai981} that repulsons, sometimes also called
``white holes'', are as generic in moduli space as their
``dual'' attractive singularities (black holes).
In addition a ``minimal'' repulson background has been analysed
in \cite{KalLin95,Gai981} using a scalar test--particle and expanding
the corresponding wave function in partial waves.
It has been shown that at the classical level no massive
scalar test--particle can reach the ``outer'' curvature
singularity at $r_h>0$.
Moreover, at the quantum level the ``inner'' singularity at the origin 
is reflecting, but the ``outer'' singularity is transparent.
In this context it has been assumed that the scalar test--particle can
tunnel through the ``outer'' curvature singularity. In addition it has
been ``suggested'' that for $r>r_h$ space--time is Minkowskian
and for $r<r_h$ space--time is Euclidean. In addition it is important
to note that the Dirac quantization condition plays a non--trivial
role in these considerations \cite{Gai981}.

It is interesting that repulson solutions are supersymmetric extensions
of the Reissner--Weyl solution \cite{Rei16}, which served as an
effective model for the electron in the 50's. Repulsons have
the ``realistic'' physical property that they
are gravitational attractive
at large distances and gravitational repulsive at short distances.
Hence, repulsons yield the usual Newtonian gravity at large
distances, but at short distances, i.e. at the order of the Planck
length $10^{-33}$ cm, for example, Newtonian gravity is not valid.

The main purpose of this article is to extend the analysis of
\cite{KalLin95,Gai981} in order to study ``non--minimal'' repulson
backgrounds and the associated antigravity effects. Although we will
restrict ourselves to a particular supersymmetric model and particular
charge configurations, most of our results are quite general.

The article is organized as follows:
To make the article sufficiently self--con\-tain\-ed we start with a short 
outline of the static solutions of the $N=2$ bosonic sector of extended 
supergravity in four dimensions. In particular we introduce the metric 
background which we call a ``repulson'' in the following. 
In the next section we define 
the specific repulson models we are investigating and outline some 
properties of the space--time they form. 
The remaining part of the paper is dedicated to the properties of these 
repulsive singularities: we model the repulson by a family of potentials 
and consider a test--particle initially moving towards the singularity 
at the origin. 
The classical analysis (section 4) tells that the particle is  reflected 
at a distance $r_{min}$ away from the origin. $r_{min}$ is 
actually bigger than $r_h$, the position of the curvature singularity 
of the underlying repulson space--time. By a Hamilton--Jacobi analysis 
the time the particle needs to get from a position $r_1$ to $r_2$ 
and the trajectory itself are calculated. 
In section 5 a quantum mechanical analysis is presented. It is defined by 
a Klein--Gordon equation on the background of the repulson. 
The geometry is spherically symmetric, so that we get an ordinary 
differential equation for the Fourier modes. For special choices 
of the parameters in the repulson model the Klein Gordon equation can 
be solved analytically. For other sets of data we present a numerical 
solution and in addition some semi--classical results in terms of a 
matched asymptotic expansion. 
In the next section we compare the numerical data with the data 
given by the matched asymptotic expansion, which has the nice feature 
that it provides the relation between the amplitude of the ingoing and the 
outgoing mode and the phase shift between them. That is, the scattering 
data can be read off. 
The asymptotic expansion of the analytical solution provides us with some 
deeper insight of the scattering behaviour even for an incoming particle 
with a low frequency. 
We finish the paper by summarizing what has been worked out, by an outline 
of the future lines of investigation and by some speculations what the 
results may imply physically.

\section{Static Solutions of N = 2 Supergravity}

The repulsons we are investigating are solutions of the bosonic sector 
of $N = 2$ supergravity in four dimensions. 
In this section we briefly outline \cite{Str95,deWLauVan85,Str90}, 
how these solutions are obtained. 

The action of the bosonic truncation of $N = 2$ supergravity in four 
dimensions includes a gravitational, vector and hypermultiplets. 
The hypermultiplets are assumed to be constant. 
The complex scalars of the vector multiplets form a sigma model the 
target space of which is special K\"{a}hler. That is, the real K\"{a}hler 
potential  $K$, which defines the sigma model metric, is entirely 
determined by a holomorphic and homogeneous function, the prepotential   
${\cal F}$. 
All the fields of the theory can be expressed in terms of the  vector 
$\gO = (X^I, {\cal F}_I)$, where $I = 0,...,N_v$ counts 
the number of the physical vectors and $N_v$ the scalars. 
The components of $\gO$ are the holomorphic sections of a line bundle 
over the moduli space. With ${\cal F}_I$ 
we denote ${\cal F}_I = \frac{\pa {\cal F}(X)}{\pa X^I}$. In order to enforce 
the configuration to be supersymmetric the so called stabilization 
equations \cite{BehLusSab98} have to be satisfied 
\[
  i \: (X^I - \bar{X}^I) \:=\: H^I(x^\mu), \qquad  
  i \: ({\cal F}_I - \bar{{\cal F}}_I) \:=\: H_I(x^\mu).  
\]
Because of the equations of motion and the Bianchi identities, the functions 
$H^I$ and $H_I$ are harmonic. 
To derive explicit solutions in general and the repulson solution in 
particular, the prepotential and the harmonic functions  
have to be specified. 
The repulsons arise in the context of a model with prepotential 
\[
  {\cal F}(S, T, U) = - STU, \qquad \mbox{with} \qquad 
  (S, T, U) = - i \: z^{1,2,3}.
\]
The $z^{1,2,3}$ are defined by so called special coordinates 
$X^0(z) = 1, \: X^A(z) = z^A, \: A = 1, ..., N_v.$
In string theory this prepotential corresponds to the classical heterotic 
$STU$--model with constant hypermultiplets. The microscopic interpretation, 
the higher order curvature corrections, the near extremal approximation,  
and the effect of quantum corrections of this class of $N = 2$ models 
has been studied extensively in \cite{Gai982,MalStrWit97,Gai983}.  

For simplicity we take all moduli to be axion--free, that is purely 
imaginary.  
Solving the stabilisation equations with respect to these constraints 
yields
\bea
 S,T,U  &=&  \frac{H^{1,2,3}}{2 X^0}, \hspace{2cm}
 X^0 \ = \ \frac{1}{2} \ \sqrt{-D/H_0}
\eea
with $D=H^1 H^2 H^3$. The harmonic functions are given by the constants 
$h$, the electric and magnetic charges $q$ and $p$, respectively
\[
  H^I(r) \:=\: h^I \: + \:  \frac{p^I}{r}, \qquad 
  H_I(r) \:=\: h_I \: + \: \frac{q_I}{r}.
\]
The charges satisfy the Dirac quantisation condition 
$p \: q \:=\: 2 \pi \: n, \: n \in \Z$. 
If we restict ourselves to static spherically symmetric solutions only,  
the most general ansatz for the metric is 
\bea
\label{metricl}
  ds^2 &=& - \frac{1}{F(r)} \ dt^2 \ + \ F(r) \ (dr^2 + r^2 d\Omega_2^2).
\eea
The metric function $F(r)$ is given by 
\[
  F \:=\: \me^{-K} \:=\: i \: ( \bar{X}^A {\cal F}_A - X^A {\cal F}_A ), 
  \qquad A = 1,2,3.
\]
Here $F^2(r)$ reads 
\[
  F^2(r) = -4 \: H_0 \: D = \sum_{n = 0}^4 \frac{\ga_n}{r^n}. 
\]
It is possible to choose the parameters of the harmonic functions such that 
the solutions correpond to   {\em repulsive} singular supersymmmetric 
states \cite{FerKalStr95}.   
It is straightforward to see whether these solutions are gravitational 
attractive or repulsive.
\begin{center}
\begin{tabular}{|c|c|c|c|c|c|}
  \hline
  $q_0$ & $p^1$ & $p^2$ & $p^3$ & (anti-)gravity & $\#$  \\
  \hline
   $+$  &  $+$  &  $+$ & $+$  & attractive     & 1  \\
  \hline
   $-$  &  $+$  &  $+$ & $+$  & repulsive      & 4  \\
  \hline
   $-$  &  $-$  & $+$ & $+$  & attractive      & 6 \\
  \hline
   $-$  &  $-$  &  $-$ & $+$  & repulsive      & 4  \\
  \hline
  $-$  &  $-$  &  $-$ &  $-$   & attractive     & 1  \\
  \hline
\end{tabular}
\end{center}
Here we denote $p,q > 0 \ (< 0)$ by $+ \ (-)$ and include possible 
permutations when denoting the number $\#$ of repulsive and attractive 
solutions. Note that any positive (negative) charge appearing in the 
solution can be interpreted as an (anti--)brane \cite{BehLusSab98,Gai981}.
Moreover, the solution with four antibranes, i.e.  all charges are 
negative, yields a {\em negative} ADM mass.  

\section{Scalar Particles in the Background of a Repulson}

In the following the repulson singularities are investigated. For that 
we follow the motion of a test--particle moving in the metric 
background produced by the repulson. The repulson gets effectively 
modelled by a family of potentials, defined by the metric function 
$F^2(r)$. The parameters $\ga_1, ... , \ga_4$ determine the shape of 
the potentials. 
 
In order to obtain flat space--time at spatial infinity $\ga_0$ 
is chosen to be equal to 1. In addition we shall consider charge 
configurations with positive ADM mass, i.e. $\ga_1 > 0$. 
For a repulson configuration the coefficient $\ga_2$ in the metric function 
is negative.  The short distance behaviour is dominated by the 
$\frac{1}{r^4}$--term in the metric function. 
Therefore the solution is gravitationally attractive at short 
distances if $\ga_4 > 0$ and gravitationally repulsive if $\ga_4 <0$. 
We restrict ourselves to the case $\ga_4 <0$ only. 
We put in a further simplification: $\ga_3$ is defined by 
\be
\label{pol3}
  \ga_1^3 + 8 \ga_3 - 4 \ga_1 \ga_2 = 0
\ee
With these restrictions on $\ga_1,...,\ga_4$ the function $F^2(r)$ 
has the generic form illustrated in {\bf figure 1a} and {\bf figure 1b}. 
{\bf Figure 1a} shows the shape of $F^2(r)$ for $\ga_3 = \ga_4 = 0$. 
The parameters $(\ga_1, \ga_2)$ are chosen to be $(10.0, -10.0),$ 
$(11.0, -9.0), (12.0, -8.0)$ and $(13.0, -7.0)$ with lines of increasing 
thickness. In {\bf figure 1b} the shape of the metric function is plotted 
for some values of $\ga_i$, 
with $\ga_3, \ga_4 \neq 0$. The triples of parameters (increasing thickness 
of the lines) are $(\ga_1, \ga_2, \ga_4) = (4.0, -1.0, -4.0),$ 
$(3.0, -1.0, -3.0), (2.0, -1.0, -2.0)$ and $(1.0, $ $-1.0, -1.0)$.  

\iffigs
\begin{figure}[bt]
\begin{minipage}[t]{6.5cm}
  \epsfxsize=6.5cm\epsfbox{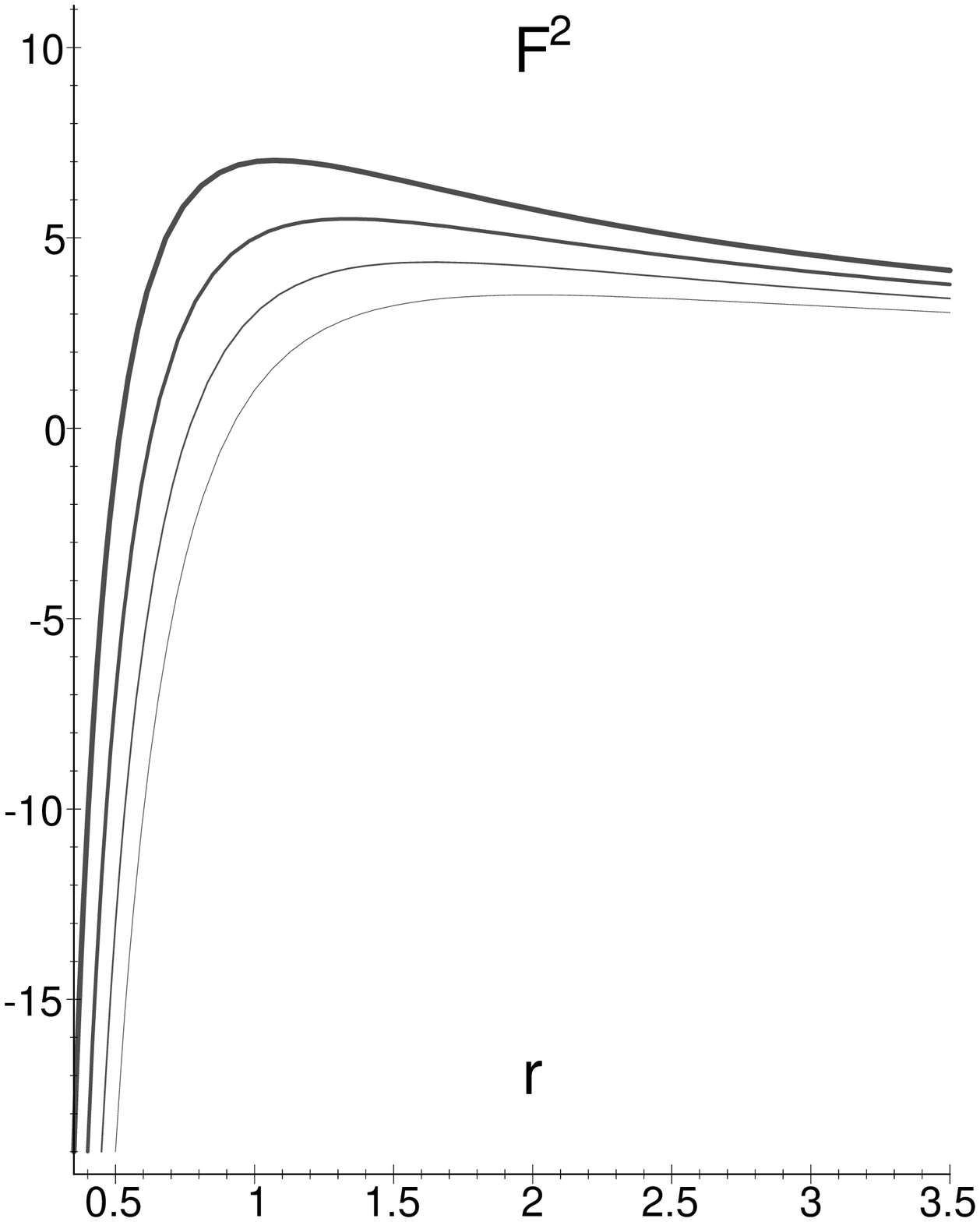}
  \small{{\bf figure 1a:} shape of the metric function $F^2(r)$ for 
    $\ga_3 = \ga_4 = 0$}
\end{minipage}
\hfill
\begin{minipage}[t]{6.5cm}
  \epsfxsize=6.5cm\epsfbox{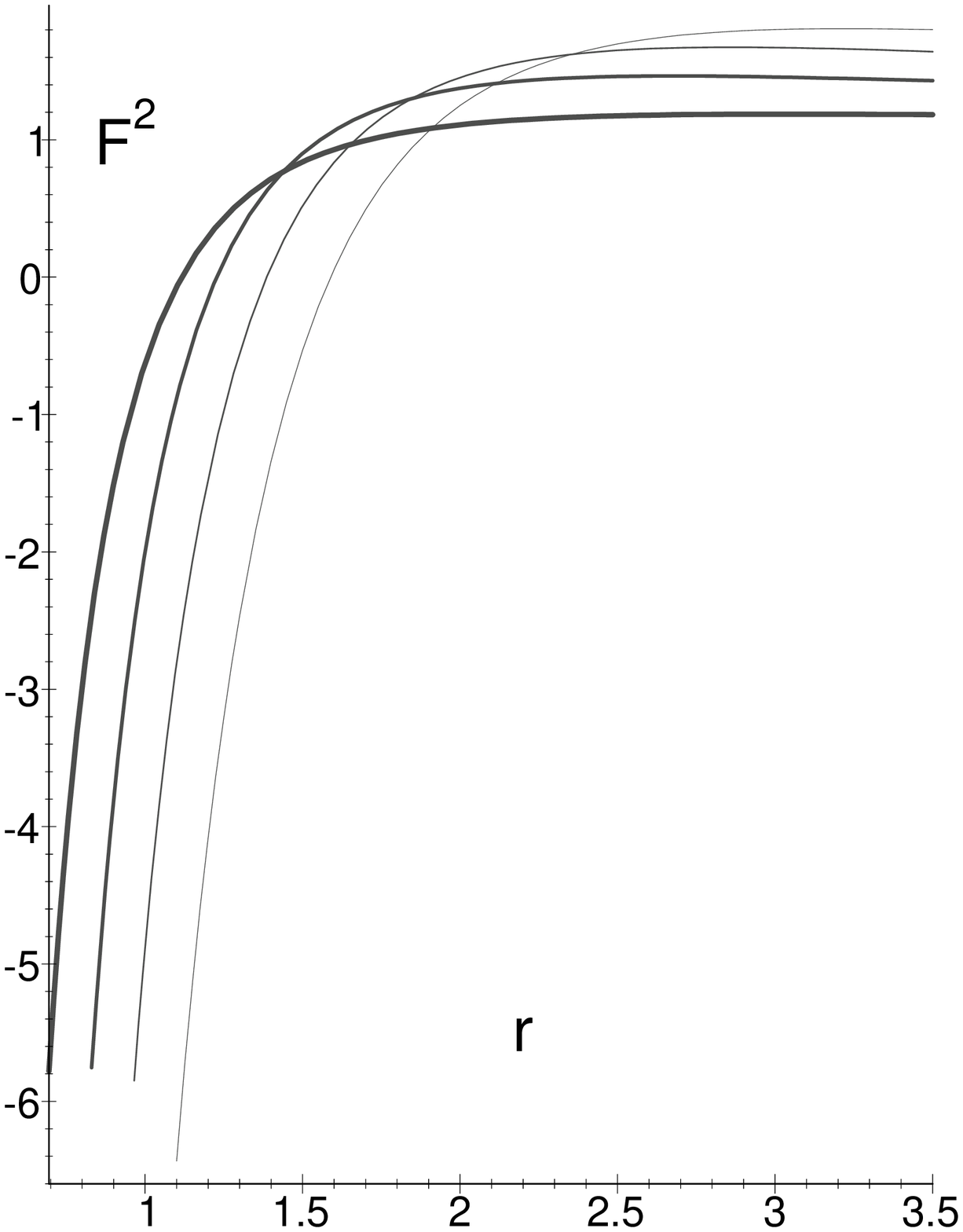}
  \small{{\bf figure 1b:} shape of the metric function $F^2(r)$ for 
    $\ga_3, \ga_4 \neq 0$}
\end{minipage}
\end{figure}
\fi

The coefficients which appear in the metric function are related to the 
physical parameters. $\ga_1 = 4 M \ge 0$, where $M$ denotes the ADM mass, 
and $\ga_2 = -4 Z^2 < 0$. Z is the central charge of the configuration. 
With (\ref{pol3}) we find $\ga_3 = -8 M(M^2 + Z^2) < 0$ and $\ga_4$ 
is a free parameter, for a repulson configuration chosen to be less 
than zero.

$F^2(r)$ has exactly one zero at a distance $r_h > 0$, which is given by 
\be
\label{nsta} 
  r_h = \sqrt{\frac{\ga_1^2}{16} + v_+} - \frac{\ga_1}{4} 
\ee
with 
\[
  v_+ = -\frac{1}{2} \left( \ga_2 - \frac{\ga_1^2}{4} \right) 
        + \sqrt{ \frac{1}{4} 
             \left( \ga_2 - \frac{\ga_1^2}{4} \right)^2 - \ga_4},   
\]
if $\ga_3$ and $\ga_4$ are not simultaneously equal to zero. 
In terms of the physical data $r_h$ reads  
\be 
  r_h = \sqrt{M^2 + v_+} - M 
\ee
with 
\[
  v_+ = 2 \: (Z^2 + M^2) + \sqrt{4 \: (Z^2 + M^2)^2 - \ga_4}.
\]
If $\ga_3$ and $\ga_4$ both vanish, the position of the zero 
is given by
\be
\label{nsts}
  r_h = \sqrt{\frac{\ga_1^2}{4} - \ga_2} - \frac{\ga_1}{2} 
      = 2 \: ( \sqrt{M^2 + Z^2} - M).  
\ee
At $r = r_h$ the metric changes its signature. 
For $r > r_h$ is is Lorentzian with the line element given by 
(\ref{metricl}). As $r \longrightarrow \infty$ the space--time is 
asymptotically Minkowskian and Schwarzschild. 
The Ricci scalar in this region is given by
\be
\label{riccis}
  R = - \frac{F^{\prime \prime}}{F^2} 
      + \frac{1}{2} \: \frac{F^{\prime 2}}{F^3} 
      - \frac{2}{r} \: {F^\prime}{F^2}. 
\ee
Approaching $r_h$ from above a curvature singularity is detected. 
In the region $0 \le r \le r_h$ the line element becomes Euclidean 
\be
\label{metrice}
  ds^2 = \frac{1}{|F|} dt^2 + |F| \left( dr^2 + r^2 d\gO^2 \right). 
\ee
The Ricci scalar is again given by the expression (\ref{riccis}). 
We find a curvature singularity when approaching $r_h$ from below. 
Ar $r=0$ all the curvature invariants remain finite. It is a coordinate 
singularity. 

\section{Classical Analysis}

In the classical limit, that is in the limit of large $r$, 
the Newtonian potential $\Phi(r)$ is given by
\bea
 \Phi(r)  &=& - \frac{1}{2} \ (g_{tt} + 1) \ = \
               - \frac{M}{r} + \frac{Z^2}{r^2}.
\eea
The corresponding strength of the gravitational field
$\Phi^\prime = \frac{M}{r^2} - \frac{2Z^2}{r^3}$ is gravitational
attractive at large distances ($r>r_c$) and gravitational repulsive
for $r<r_c$. The critical distance where gravitational repulsion and
attraction yield a vanishing net force is given by $r_c = 2 Z^2/M$. 
For the classical limit still to be valid at $r_c$ we derive a constraint 
on the central charge $Z$ and the ADM mass $M$. We got the Newtonian 
potential with the requirement $| \frac{4 M}{r} - \frac{4 Z^2}{r^2} | < 1$. 
This must hold at $r = r_c$ in particular, and it follows $Z^2 > M^2$.   
For massless repulsons the Newtonian potential is always repulsive.

To consider the motion of a classical scalar test--particle we choose 
a suitable plane with e.g., $\theta=\pi/2$. The corresponding
trajectory of the test--particle of mass $m$ in this plane can be 
determined using Hamilton--Jacobi theory (see e.g., \cite{KalLin95, 
LanLif75}).  In the classical limit for the repulson background 
the Hamilton--Jacobi equation 
\bea
 g^{\mu\nu} \frac{\pa \gC}{\pa x^\mu}
            \frac{\pa \gC}{\pa x^\nu}  \ + \ m^2 &=& 0
\eea
becomes 
\bea
  F \left (\frac{\pa \gC}{\pa t} \right )^2
  - \frac{1}{F} \left (\frac{\pa \gC}{\pa r} \right )^2
  - \frac{1}{r^2 F} \left (\frac{\pa \gC}{\pa \phi} \right)^2
  - m^2 &=& 0.
\eea
By the general procedure for solving the Hamilton--Jacobi equation we
take $\gC$ to be of the form
\bea
  \gC  &=& -Et \ + \ L \phi \ + \ \gC_r(r), 
\eea
with energy $E$ and angular momentum $L$. This yields 
\bea
  \gC_r(r)   &=& \int \ dr \
  \sqrt{E^2 F^2 - \frac{L^2}{r^2} - m^2 F}.
\eea
Thus, from $\frac{\pa \gC}{\pa E}=0$ it follows that a test--particle
takes the following time to move from $r_1$ to $r_2$
\bea
  t  &=& E \: \int_{r_1}^{r_2} \ dr \
  \frac{F^2}{\sqrt{E^2 F^2 - \frac{L^2}{r^2} - m^2 F}}
\eea
For $L=0$ a massive test--particle becomes reflected
by the repulson at
\bea
 r_{min}  &=& \frac{2}{\epsilon}
   \left ( \sqrt{M^2 + \epsilon Z^2} - M \right ) \ > \ r_h,
 \hspace{1cm} \epsilon = 1 - \frac{m^4}{E^4}. 
\eea
Moreover, for $L \neq 0$ a massless test--particle becomes reflected
by the repulson at
\bea
 r_{min}  &=& 2
 \left (
  \sqrt{M^2 + Z^2 + \frac{L^2}{4E^2} } - M
 \right ) \ > \ r_h.
\eea
The trajectory itself is determined by $\frac{\pa \gC}{\pa L}=0$, i.e.
\bea
 \phi  &=& \int_{r_1}^{r_2} \ dr \
 \frac{-L}{r^2 \sqrt{E^2 F^2 - \frac{L^2}{r^2} - m^2 F}}.
\eea


\section{Quantum Mechanical Analysis}

For the quantum mechanical analysis we consider a massless scalar 
test--particle with wave function $\tilde{\psi}$ satisfying the 
Klein--Gordon equation in the background of the repulson:
\bea
\label{KG}
 \pa_\mu \ \left ( \sqrt{-g} \ g^{\mu \nu} \ \pa_\nu \ \tilde{\psi} \right )
   &=& 0.
\eea
Writing $\tilde{\psi} = \Psi(t,r) \: Y_{lm}(\gt, \phi)$ we obtain the 
same equation with the metric in either the Lorentzian (\ref{metricl})
 or the Euclidean region (\ref{metrice}) 
\be
\label{pde}
  \gD \Psi - \nu \: |F^2| \: \Psi_{tt} = 0,
\ee
where $\gD$ is the flat space Laplace operator in spherical coordinates
\be
\label{la}
  \gD = \pa_r^2 + \frac{2}{r} \pa_r - \frac{l(l+1)}{r^2},  
\ee
and $\nu$ is $1$ for $r \ge r_h$ and $-1$ for $r < r_h$. 
Since (\ref{pde}) is a linear partial differential equation with 
$t$--independent coefficients we may perform a Fourier transform 
with respect to $t$, that is $\Psi = \me^{-i \go t} \: \psi(r)$, to obtain 
\be
\label{ode}
  \gD \psi + \nu \: \go^2 \: |F^2| \: \psi = 0.
\ee
This is a Schr{\"o}dinger equation 
\be
\label{schroe}
  \gD_r \psi + (E - V(r)) \: \psi = 0, 
\ee
with 
\bea
  E &=&  \go^2, \nonumber \\
  V(r) &=& -  \left( 
      \frac{c_1}{r} + \frac{c_2}{r^2} + \frac{c_3}{r^3} + \frac{c_4}{r^4} 
    \right), \nonumber \\
  \gD_r &=& \pa_r^2 \: + \+ \frac{2}{r} \: \pa_r.
\eea
The coefficients in terms of the physical data are given by
\[
  c_1 = 4 M \go^2, \quad c_2 = - 4 Z^2 \go^2 - l(l+1), 
  \quad c_3 = -8 M \go^2 \: (Z^2 + M^2), \quad c_4 = \ga_4 \go^2. 
\]
For $l = 0$ the potential is, up to a factor $- \go^2$, equal to the 
metric function $F^2$, so that the shape of the potential can be 
read off {\bf figure 1a} and {\bf figure 1b}, respectively.
Let us first study the special situation $\ga_3 = \ga_4 = 0$ \cite{Gai981}. 
In this case the differential equations  (\ref{schroe}) simplify to 
\be
\label{schroes}
  \gD_r \psi + (E - V(r)) \: \psi = 0, 
\ee
with 
\bea
  E &=& \go^2 \nonumber \\
  V(r) &=& -  \left( \frac{c_1}{r} + \frac{c_2}{r^2}  \right).  
\eea  
Here we have taken into account that at the border of the Euclidean and Minkowski 
region of the metric not only $\nu$ but $F^2(r)$ changes its sign too.  
We now solve the differential equation. 

The transformation to a new coordinate 
$\rho = 2 i \go \: r$ brings the differential 
equation into the form 
\[
  \pa_\rho^2 \psi + \frac{2}{\rho} \: \pa_\rho \psi + 
  \left(
     - \frac{1}{4} + \frac{n}{\rho} - \frac{s(s+1)}{\rho^2} 
  \right) \psi = 0,
\]
where $n = -i \: c_1/(2 \go)$ and $-s(s+1) = c_2$. 
The Ansatz $\psi(\rho) = ( i \rho)^s \me^{-\frac{\rho}{2}} f(\rho)$ 
yields the Kummer equation for $f$
\[
  \rho \: \pa_\rho^2 f + (2s + 2 - \rho) \: \pa_\rho f + (n-s-1) \: f = 0,
\]
so that the wave function turns out to be 
\be
\label{wa1} 
  \psi(r) = (2 \go r)^s \me^{-i \go r} \times 
\ee
\[
  \left[
    C_1 \: \F(1+s-n, 2s+2, 2 i \go r) 
    + C_2 \: (2 i \go r)^{-(1 + 2s)} \F(-s-n, -2s, 2 i \go r)  
  \right],
\]
where we have choosen $s \geq - \frac{1}{2}$. 
The functions $F$ are the confluent hypergeometric functions. 
The second term in the square brackets of equation (\ref{wa1}) 
is singular at $r = 0$.  
If we require $\psi(r)$ to be regular at the origin, $C_2$ has to be zero. 

\iffigs
\begin{figure}[bt]
\begin{minipage}[t]{6.5cm}
  \epsfxsize=6.5cm\epsfbox{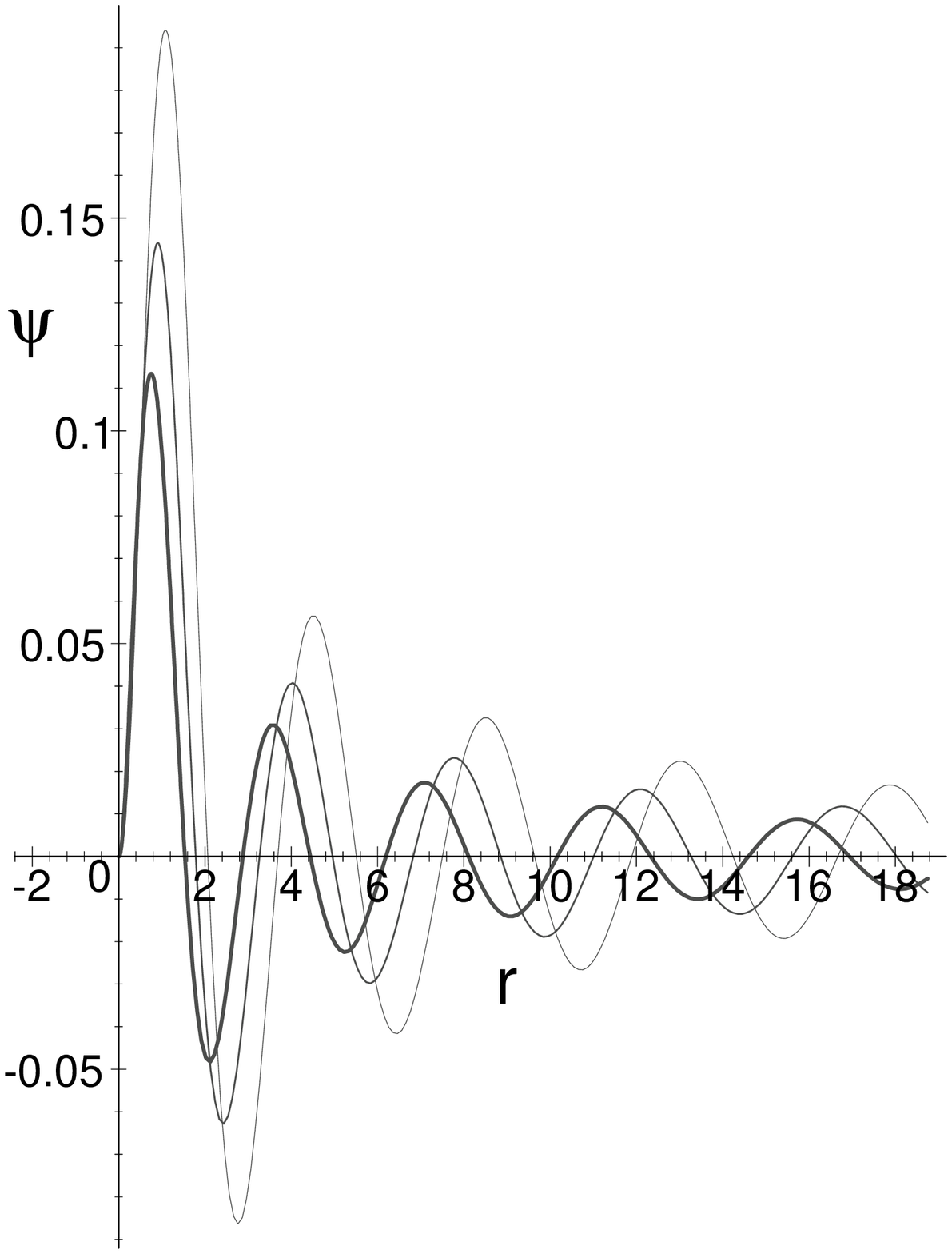}
  \small{{\bf figure 2a:} Wave functions for \\ $\ga_3 = \ga_4 = 0$.}
\end{minipage}
\hfill
\begin{minipage}[t]{6.5cm}
  \epsfxsize=6.5cm\epsfbox{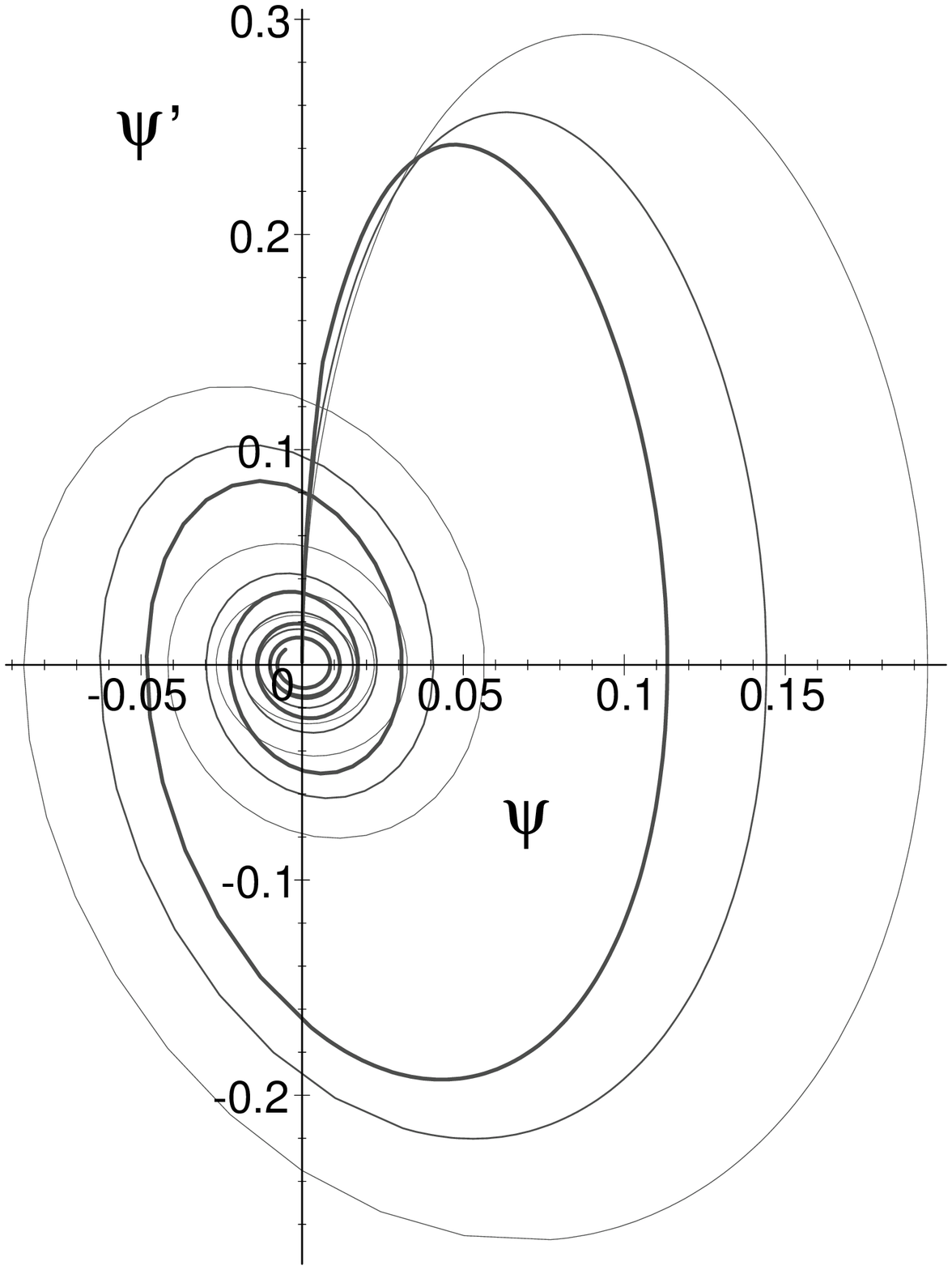}
  \small{{\bf figure 2b:} Phase space diagram for \\ $\ga_3 = \ga_4 = 0$.}
\end{minipage}
\end{figure}
\fi

The solutions are illustrated in {\bf figure 2a} and {\bf figure 2b}. 
In {\bf figure 2a} the wave functions are plotted for   
$l = 0$ and $\go = 1$. The pairs of parameter values 
$(\ga_1, \ga_2) = (11.0, -9.0), \: (12.0, -8.0), \: (13.0, -7.0)$   
are indicated by lines of increasing thickness.  
In {\bf figure 2b} we see the phase space diagram for the wavefunctions 
with parameter values as indicated above.   

If $\ga_3$ and $\ga_4$ are not zero we have to integrate 
the differential equation (\ref{schroe}) numerically. 
The wave functions and the phase space diagrams are shown in 
{\bf figure 3a} and {\bf figure 3b}.
  
\iffigs
\begin{figure}[bt]
\begin{minipage}[t]{6.5cm}
  \epsfxsize=6.5cm\epsfbox{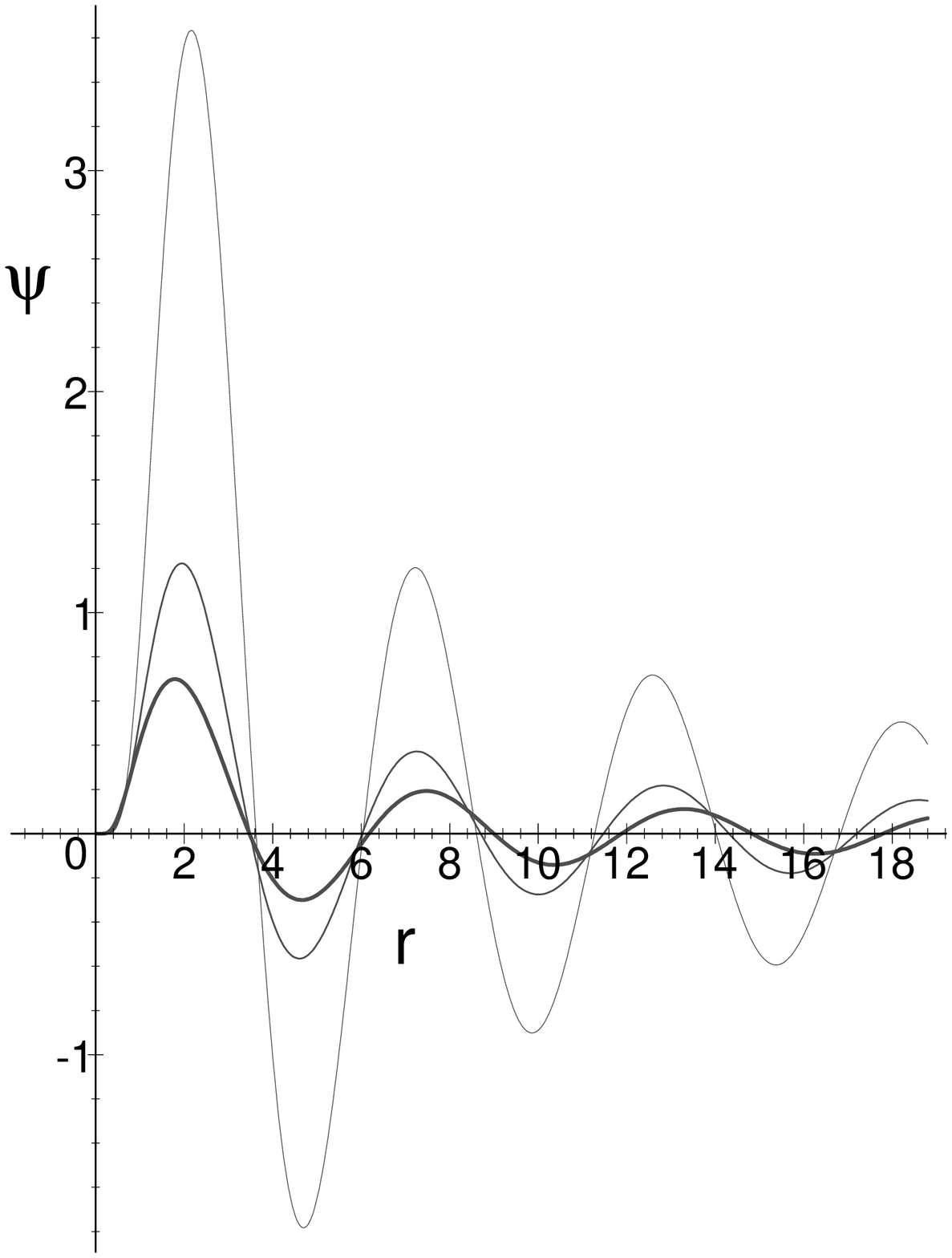}
  \small{{\bf figure 3a:} Wave functions for $\ga_3, \ga_4 \neq 0$. }
\end{minipage}
\hfill
\begin{minipage}[t]{6.5cm}
  \epsfxsize=6.5cm\epsfbox{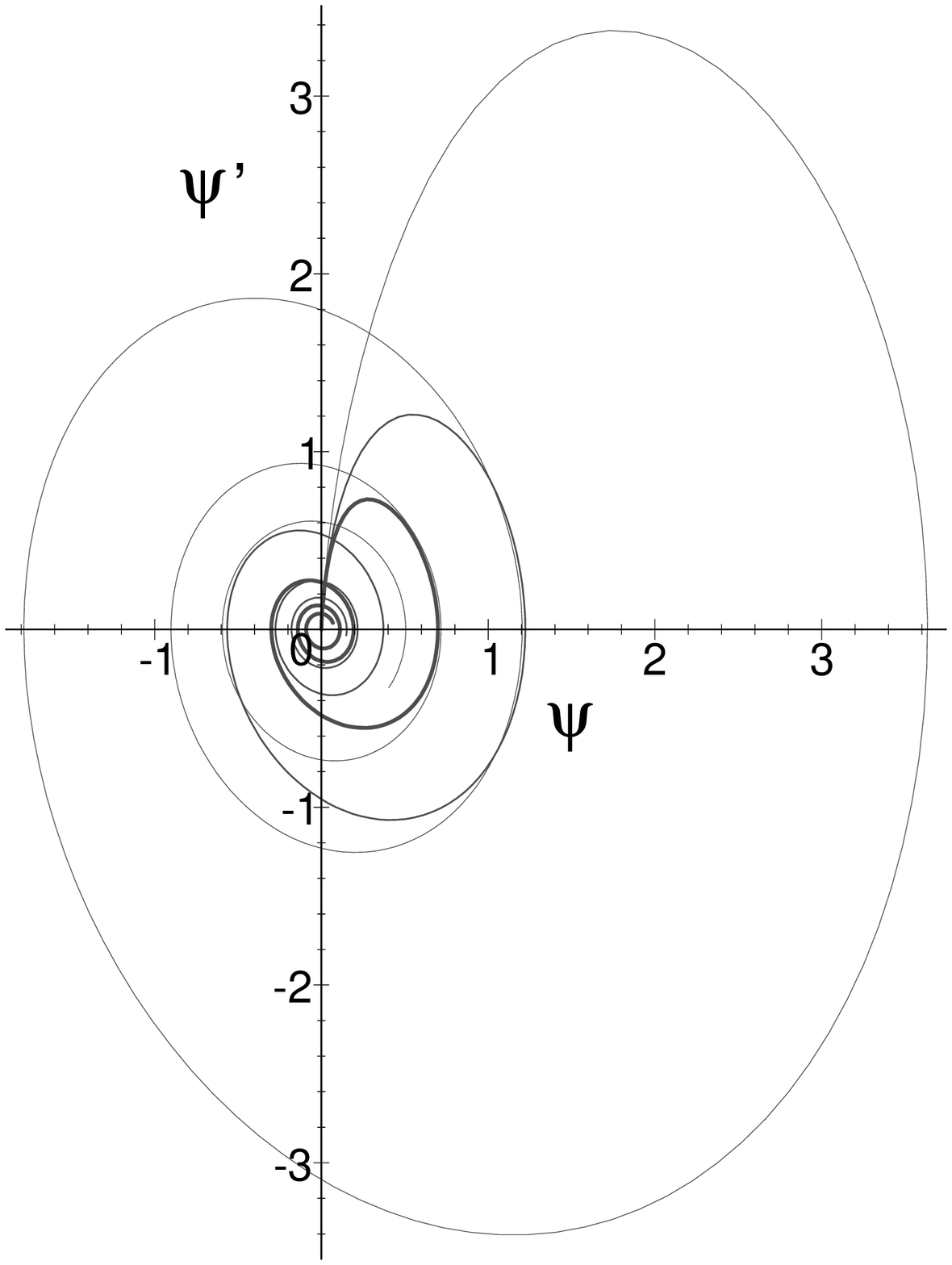}
  \small{{\bf figure 3b:} Phase space diagram for $\ga_3,$ $\ga_4 \neq 0$. }
\end{minipage}
\end{figure}
\fi

Again we have choosen $l$ to be equal to zero and $\go = 1.0$. 
The lines of increasing thickness correspond to the following triples 
of parameter values $(\ga_1, \ga_2, \ga_4) = (4.0, -1.0, -4.0), \: 
(3.0, -1.0, -3.0), \: (2.0, -1.0, 2.0)$ ($\ga_3$ is determined by equation 
(\ref{pol3})). 

In {\bf figure 4a, 4b, 5a} and {\bf 5b} we show how the shape of 
the wave function and the phase space diagrams change with increasing 
or decreasing value of $\go$. 
In these cases the parameters are $\ga_1 = 1.0, \ga_2 = -1.0$ and 
$\ga_4 = -1.0$.

\iffigs
\begin{figure}[bt]
\begin{minipage}[t]{6.5cm}
  \epsfxsize=6.5cm\epsfbox{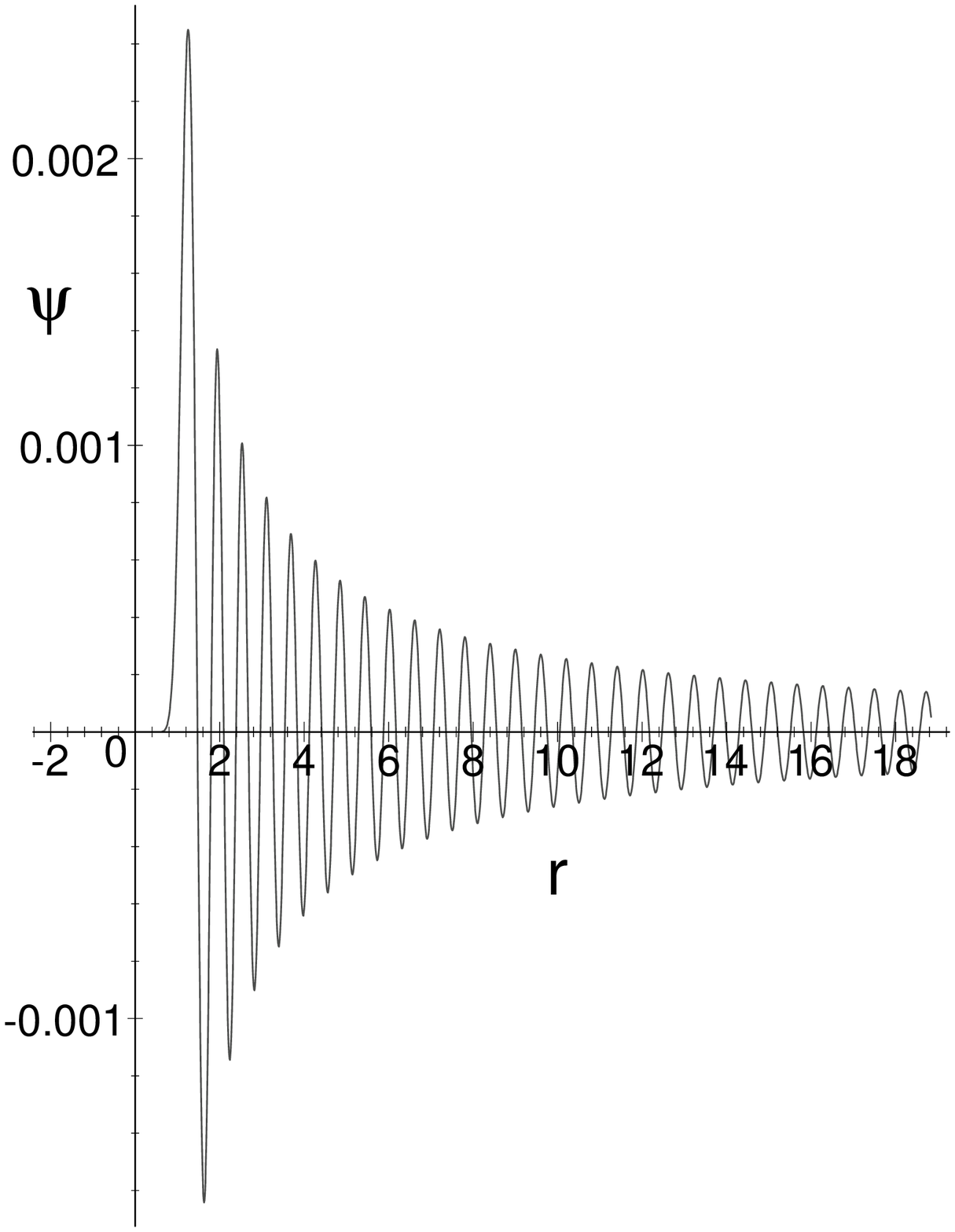}
  \small{{\bf figure 4a:} Wave function for $\go = 10$. }
\end{minipage}
\hfill
\begin{minipage}[t]{6.5cm}
  \epsfxsize=6.5cm\epsfbox{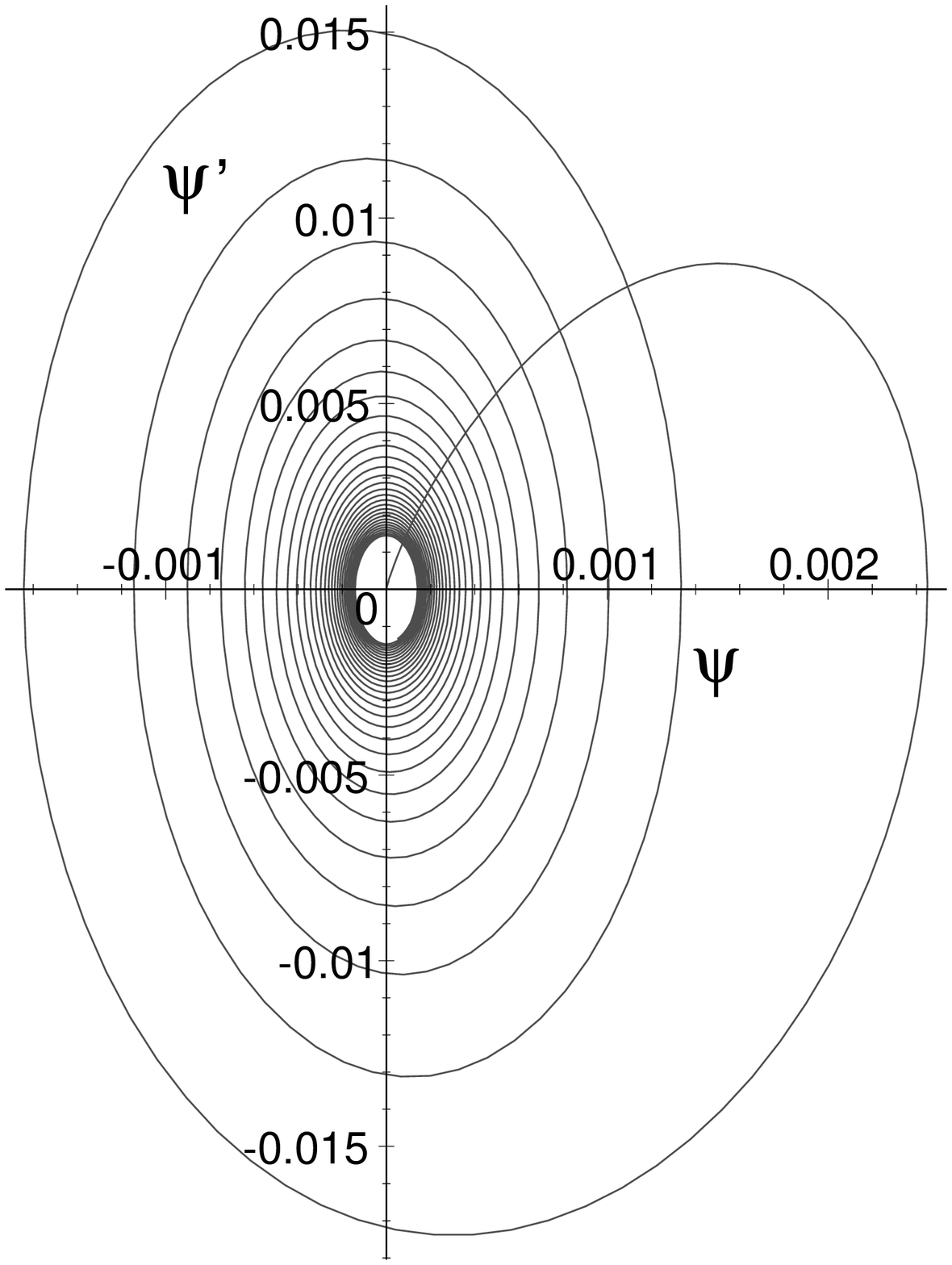}
  \small{{\bf figure 4b:} Phase space diagram for \\ $\go = 10$.}
\end{minipage}
\end{figure}
\fi

\iffigs
\begin{figure}[bt]
\begin{minipage}[t]{6.5cm}
  \epsfxsize=6.5cm\epsfbox{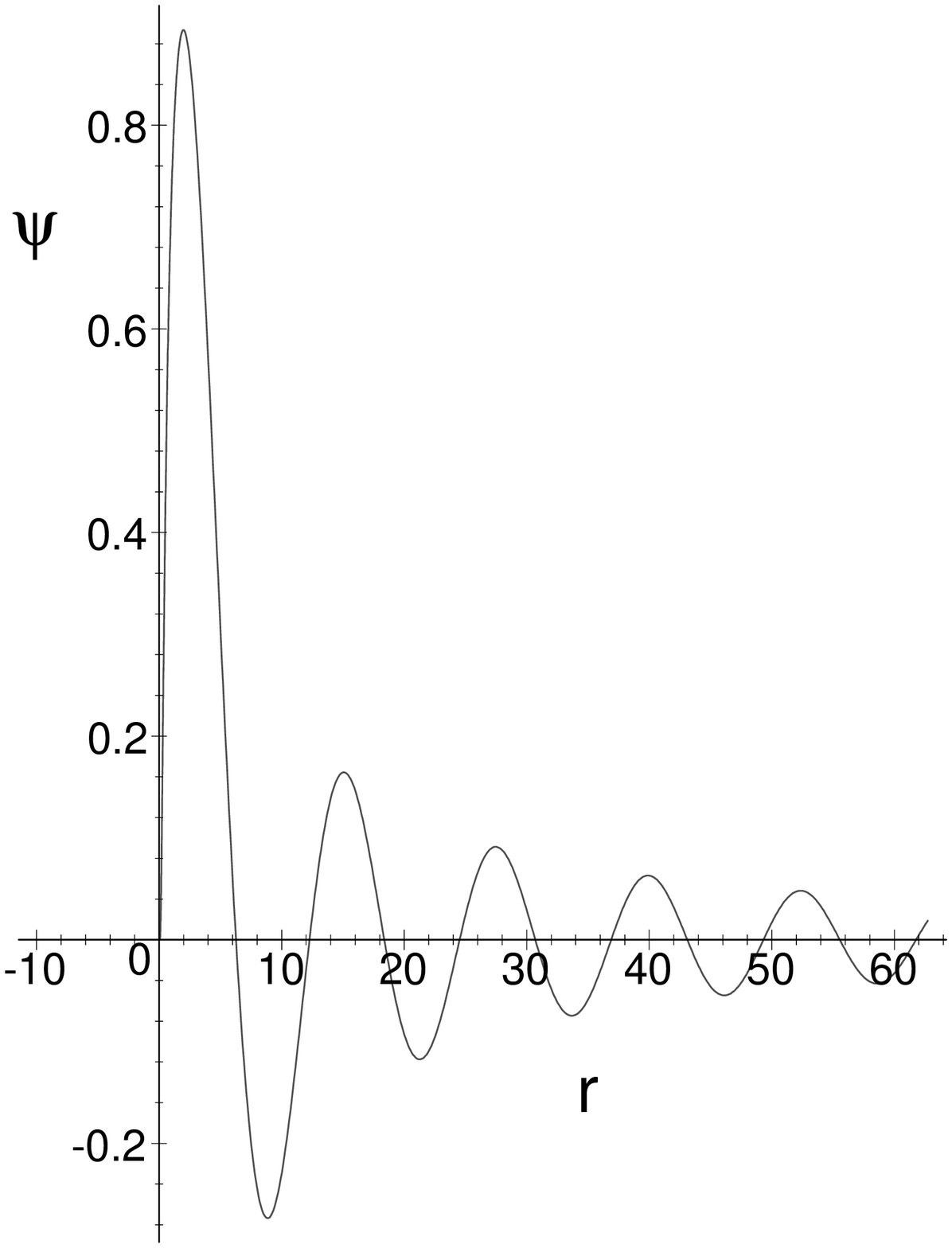}
  \small{{\bf figure 5a:} Wave function for $\go = 0.5$. }
\end{minipage}
\hfill
\begin{minipage}[t]{6.5cm}
  \epsfxsize=6.5cm\epsfbox{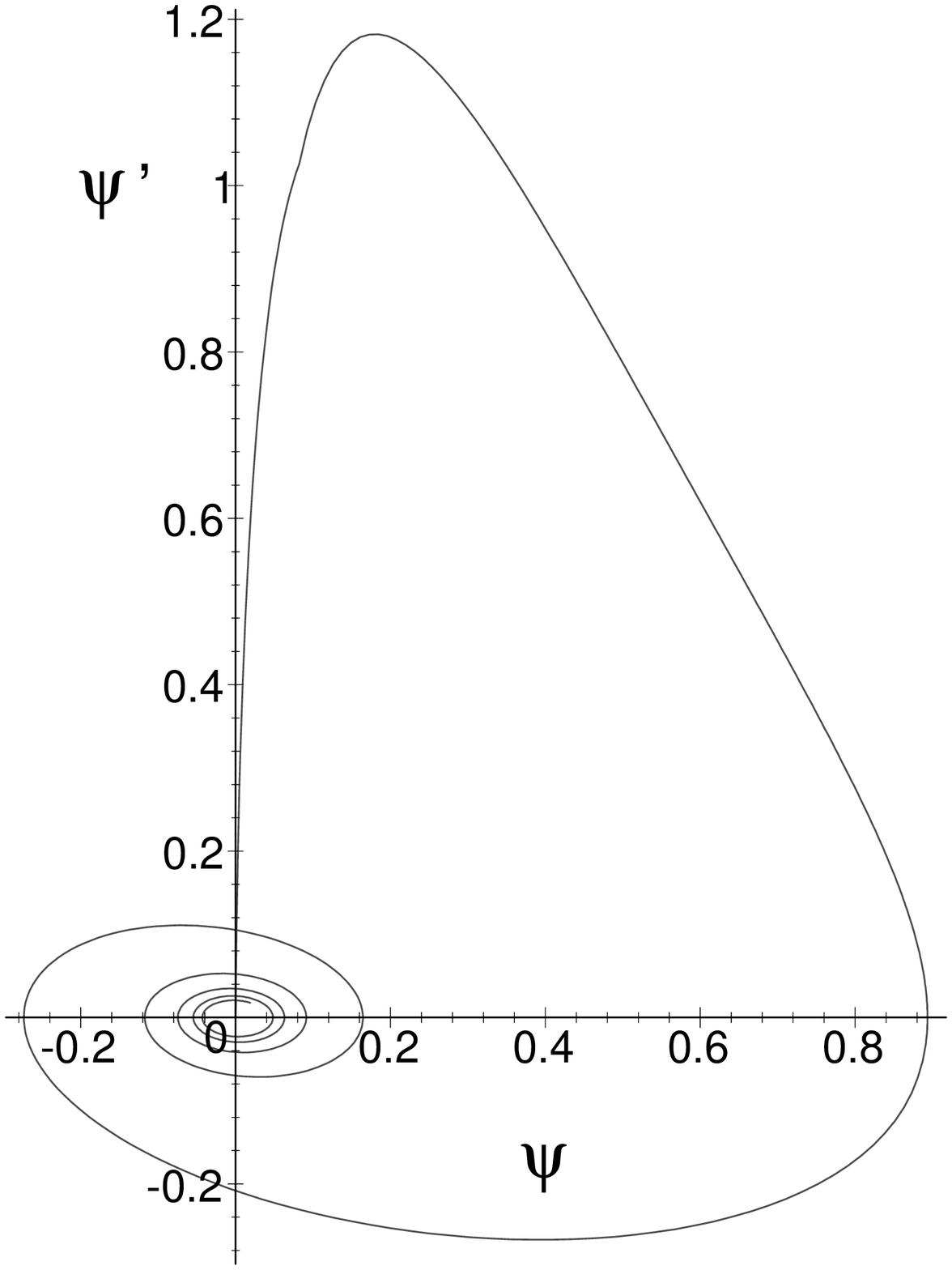}
  \small{{\bf figure 5b:} Phase space diagram for \\ $\go = 0.5$.}
\end{minipage}
\end{figure}
\fi

Some information about the scattering can be obtained analytically 
in terms of a WKB approximation \cite{BenOrs78}. 
Applying the transformation $\psi(r) \: = \: \gt(r)/r$ to (\ref{schroe}), 
we obtain a differential equation for $\gt(r)$  
\be
\label{wkbdgl}
  \pa_r^2 \gt \: - \: Q(r) \: \gt \: = \: 0,   
\ee
with 
\[
  Q(r) \: = \: \go^2 \: Q_0(r) \:+\: Q_2(r) 
       \: = \:  - \go^2 \: F^2(r) \:+\:  \frac{l(l+1)}{r^2}.  
\]
If we set $\gep = 1/\go$ and substitute the WKB ansatz  
\[
  \gt(r) = \exp \left( \frac{1}{\gep} \: 
    \left[\sum_{n = 0}^{\infty} \gep^n \: S_n(r) \right]
                \right) 
\]
into equation (\ref{wkbdgl}) differential equations for the 
$S_n$'s are obtained. If the series is truncated 
after $S_1$ what results is called the physical optics 
approximation. 
The corresponding differential equations are 
\bsea 
\label{wkbs}
  && S_0^{\prime 2} \: = \: Q_0,  \\
  && 2 \: S_0^\prime S_1^\prime + S_0^{\prime \prime}  \: = \: 0, \\
  && 2 \: S_0^\prime S_2^\prime + S_1^{\prime \prime} + S_1^{\prime 2} 
    \:=\: Q_2. 
\esea
The solutions are 
\bsea
  S_0^\pm(r) &=& \pm \int_{r_h}^r \: \sqrt{Q_0(t)} \: dt, \\ 
  S_1(r) &=& -\frac{1}{4} \: \ln |Q_0(r)|, \\
  S_2(r) &=& \pm \: \int_{r_h}^r 
    \left(
      \frac{Q_0^{\prime \prime}}{8 Q_0^{\frac{3}{2}}}  
        - \frac{5}{32} \frac{Q_0^{\prime 2}}{Q_0^{\frac{5}{2}}}  
    \right) \: dt, 
    \qquad \mbox{for} \qquad l \:=\: 0. 
\esea
Therefore the physical optics approximation is  
\[
  \gt(r) = \frac{1}{\sqrt[4]{|Q_0(r)|} } \: 
  \left[
    C_1 \: \me^{ \frac{1}{\gep} \int_{r_h}^r \sqrt{Q_0(t)} \: dt } \:+\:  
    C_2 \: \me^{ -\frac{1}{\gep} \int_{r_h}^r \sqrt{Q_0(t)} \: dt }
  \right]. 
\]
It should be noted that the angular momentum component of the potential 
does not contribute to the physical optics approximation. 

At $r = r_h$ the potential $Q_0(r)$ vanishes and the WKB approximation 
is not valid. 
Therefore we need to split the interval $[0,\infty)$ into three regions, 
a region I near $r = 0$, a region II near $r = r_h$ and a region III far 
outside, and they need to be investigated separately. 

For $r < r_h$ the potential $Q_0(r)$ is positive and in addition 
we require $\psi(0)$ not to blow up exponentially. Consequently the wave 
function $\psi_I(r)$ in this region is given by 
\[
  \psi_I(r) = \frac{C_1}{r \: \sqrt[4]{|Q_0(r)|}} 
    \: \me^{\frac{1}{\gep} \int_{r_h}^r \sqrt{Q_0(r)} \: dt} 
    \qquad \mbox{for} \qquad r < r_h. 
\]
As mentioned above the WKB approximation does not hold near $r = r_h$. 
However the potential near $r_h$ can be approximated by a linear  
function with negative slope so that equation (\ref{wkbdgl}) in region 
II becomes approximately 
\[
  \gt^{\prime \prime} = \frac{a}{\gep^2} \: (r - r_h) \: \gt(r), 
  \quad \mbox{where} \qquad
  a = Q_0^\prime(r_h) < 0.
\]
The solution of this approximate equation is  
\[
  \psi_{II}(r) = \frac{1}{r} \: 
  \left\{
    C_a \: \Ai\left[ \gep^{-\frac{2}{3}} \sqrt[3]{-a} \: (r_h -r ) \right] 
    \: + \:  
    C_b \: \Bi\left[ \gep^{-\frac{2}{3}} \sqrt[3]{-a} \: (r_h -r) \right]
  \right\}.
\]
In region III $Q_0(r)$ is negative, so that $S_0(r)$ is pure imaginary 
and the wave function $\psi_{III}(r)$ can be written as 
\[
  \psi_{III}(r) = \frac{1}{r \: \sqrt[4]{|Q_0(r)|} } \: 
  \left\{
    C_o \: \me^{\frac{i}{\gep} \: \int_{r_h}^r \sqrt{ |Q_0(t)| } \: dt } 
    \: + \:  
    C_i \: \me^{- \frac{i}{\gep} \: \int_{r_h}^r \sqrt{ |Q_0(t)| } \: dt } 
  \right\}.
\]
We now try to patch these approximations to gether to obtain a solution 
on $[0, \infty)$. 
The matching procedure is done 
by estimating the range of validity of the solution in each region. 
In order that the WKB approximation be valid on an interval it is 
necessary that 
\be
\label{wkberr}
  \frac{S_0}{\gep} \: \gg \: S_1 \: \gg \: \gep S_2 \:\gg\: ... \: \gg 
  \gep_{n-1} S_n(x) \qquad \mbox{for} \qquad \gep \longrightarrow 0.
\ee
We shall come back to this point in section 6. 

In the overlapping regions we take into account the
asymptotic behaviour of the respective solutions.

To match the functions $\psi_I(r)$ and $\psi_{II}(r)$ the asymptotic 
behaviour of the wave functions have to agree. 
For large positive arguments the Airy functions behave like 
\[
  \Ai(t) \: \sim \: \frac{1}{2 \sqrt{\pi}} \: t^{-\frac{1}{4}} \: 
    \me^{- \frac{2}{3} t^{\frac{3}{2}}} , \qquad 
  \Bi(t) \: \sim \: \frac{1}{\sqrt{\pi}} \: t^{-\frac{1}{4}} \: 
    \me^{ \frac{2}{3} t^{\frac{3}{2}} }. 
\]
This shows that $C_b = 0$ and 
$C_a = \frac{2 \sqrt{\pi}}{(-a \gep)^{\frac{1}{6}} } \: C_1$. 
We now fit the wave function in region II to region III. 
For large negative arguments the Airy function $\Ai$ 
behaves like    
\[
  \Ai\left( \gep^{-\frac{2}{3}} \sqrt[3]{-a} \: (r_h - r) \right) \: \sim \:  
  \frac{\gep^{\frac{1}{6}}}{\sqrt{\pi}}  \: \frac{ \sin \left( 
    \frac{2}{3 \gep} \sqrt{-a} \: (r - r_h)^{\frac{3}{2}} + \frac{\pi}{4} 
              \right)}{(-a)^{\frac{1}{12}} \: 
                         (r - r_h)^{\frac{1}{4}} }, 
\]
that is 
\[
  \psi_{II}(r) \: \sim \: 
    C_a \: \frac{\gep^{\frac{1}{6}} }{\sqrt{\pi} \: r} \: \frac{ \sin \left( 
    \frac{2}{3 \gep} \sqrt{-a} \: (r - r_h)^{\frac{3}{2}} + \frac{\pi}{4} 
              \right)}{(-a)^{\frac{1}{12}} \: (r - r_h)^{\frac{1}{4}} }.
\]
In the overlapping region $\psi_{III}(r)$ is approximated by 
\[
  \psi_{III}(r) \sim \frac{1}{ r \: \sqrt[4]{-a} \: \sqrt[4]{r - r_h} } \: 
  \left[
    C_o \: \me^{\frac{2}{3 \gep}\: i \: \sqrt{-a} \: (r - r_h)^{\frac{3}{2}}} 
    \: + \: 
    C_i \: \me^{- \frac{2}{3 \gep}\: i \: \sqrt{-a} \: (r - r_h)^{\frac{3}{2}}}  
  \right]
\]
The matching determines $C_o$ and $C_i$ via  
\[
  \frac{C_a \: (-a \gep)^{\frac{1}{6}}}{\sqrt{\pi}}  \: 
  \sin (x + \frac{\pi}{4}) 
 \: = \: 
    C_o \: \me^{i \: x}  \: + \: C_i \: \me^{- i \: x}, 
\]
with $x = \frac{2}{3 \gep} \sqrt{-a} \: (r - r_h)^{\frac{3}{2}}$.  
It follows that 
\[
   C_o = C_a  \: \frac{(-a \gep)^{\frac{1}{6}}}{ 2 i \: \sqrt{\pi}} \: 
       \me^{\frac{1}{4} \: i\: \pi}, \qquad   
   C_i = - C_a  \: \frac{(-a \gep)^{\frac{1}{6}}}{ 2 i \: \sqrt{\pi}} \: 
       \me^{-\frac{1}{4} \: i\: \pi}   
\]
and therefore 
\[
  \psi_{III}(r) =  C_a \: \frac{(-a \gep)^{\frac{1}{6}}}{\sqrt{\pi}}
     \frac{1}{r \: \sqrt[4]{|Q_0(r)|}} \: 
  \sin \left( \frac{1}{\gep} 
    \int_{r_h}^r \sqrt{|Q_0(r)|} \: dt + \frac{\pi}{4} 
       \right).
\]
Thus we have obtained the approximations ($\gep = \frac{1}{\go}$): 
\bsea 
\label{wkb123}
  \psi_I(r) &=& C_a \: \frac{(-a \gep)^{\frac{1}{6}}}{2 \sqrt{\pi}} 
   \frac{1}{r \: \sqrt[4]{|Q_0(r)|} } \: \me^{\frac{1}{\gep} \int_{r_h}^r 
  \sqrt{|Q_0(t)|} \: dt},  \\
  \psi_{II}(r) &=&  C_a \: 
  \frac{\Ai \left( \gep^{-\frac{2}{3}} \sqrt[3]{-a} \: (r_h -r) \right)}{r}, 
    \\
  \psi_{III}(r) &=&  C_a \: \frac{(-a \gep)^{\frac{1}{6}}}{\sqrt{\pi}}
     \frac{1}{r \: \sqrt[4]{|Q_0(r)|}} \: 
  \sin \left( \frac{1}{\gep} 
    \int_{r_h}^r \sqrt{|Q_0(r)|} \: dt + \frac{\pi}{4} 
       \right).
\esea


\section{Discussion}

A comparison of the numerical solution and the physical optics 
approximation is given in {\bf figure 6a} and {\bf figure 6b}. 

\iffigs
\begin{figure}[bt]
\begin{minipage}[t]{6.5cm}
  \epsfxsize=6.5cm\epsfbox{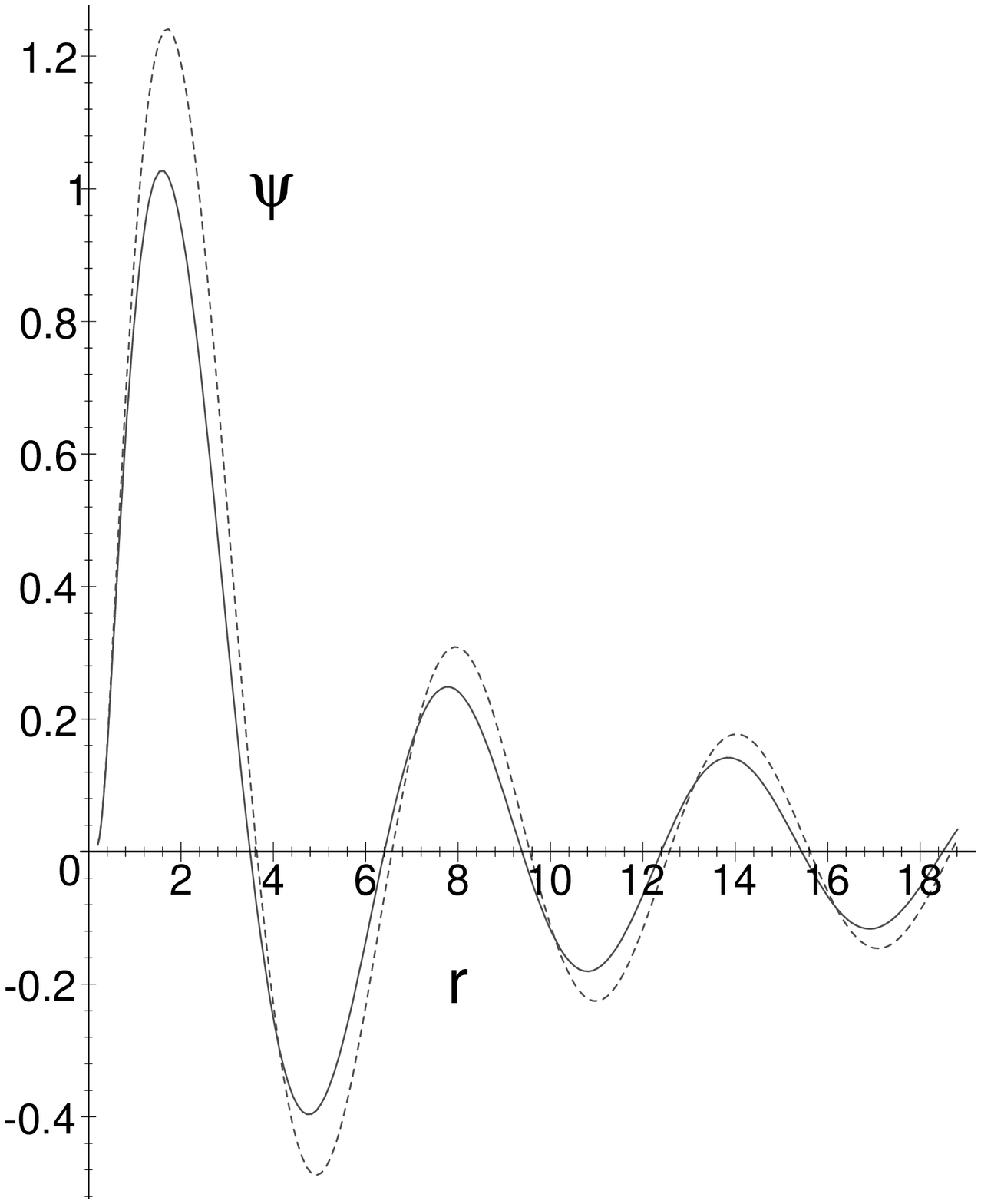}
  \small{{\bf figure 6a:}  Numerical (dashed line) and WKB 
  solution (solid line) for $\go = 1.0$}
\end{minipage}
\begin{minipage}[t]{6.5cm}
  \epsfxsize=6.5cm\epsfbox{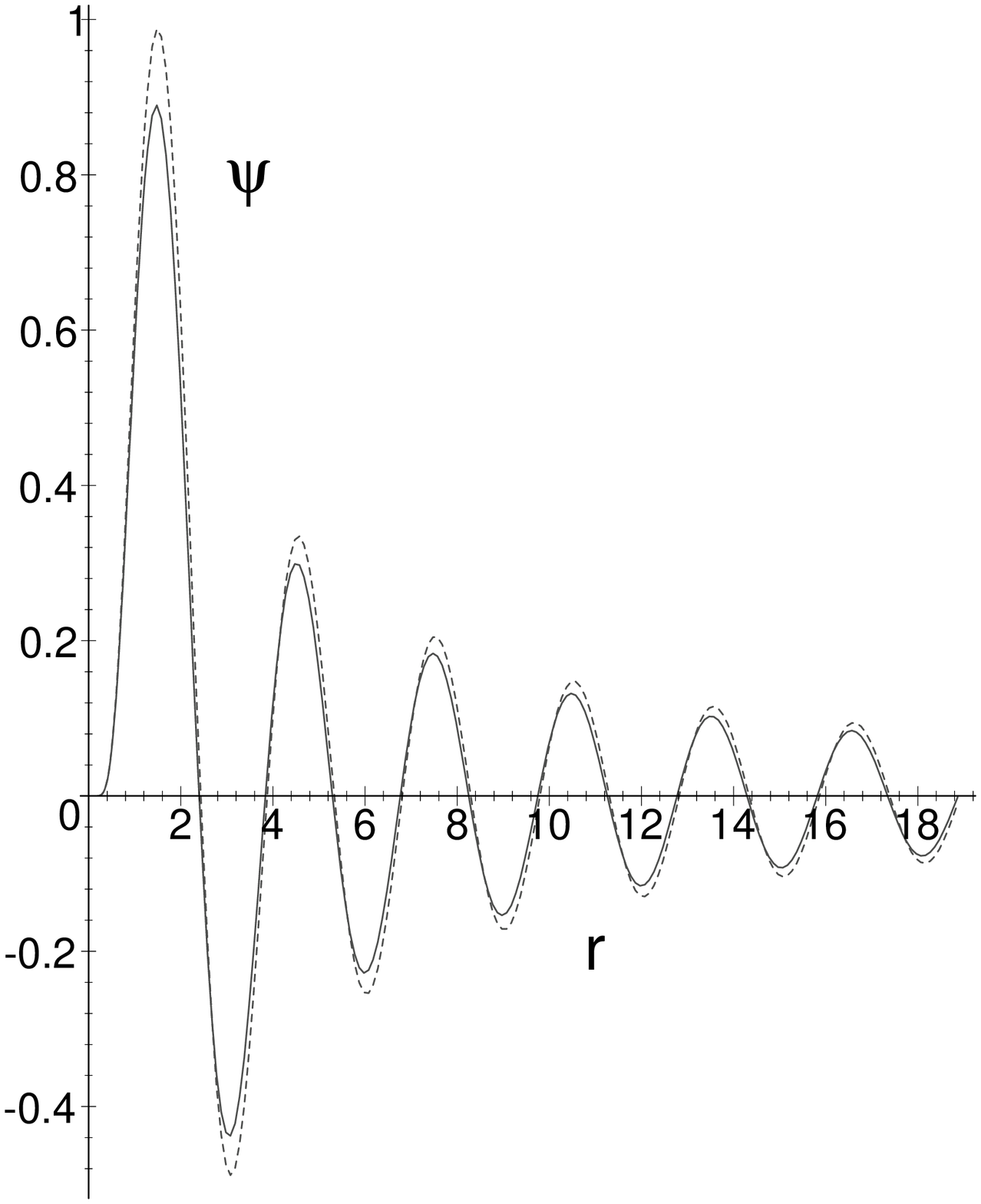}
  \small{{\bf figure 6b:} Numerical (dashed line) and WKB 
  solution (solid line) for $\go = 2.0$}
\end{minipage}
\end{figure}
\fi

{\bf Figure 6a} shows the numerical and the WKB solution for 
the parameter values $l = 0.0, \go = 1.0, \ga_1 = 1.0, 
\ga_2 = -1.0$ and $\ga_4 = - 1.0$.  
{\bf Figure 6b} is a plot of the numerical solution and the WKB solution 
for a larger value, $\go = 2.0$. 
The normalizations of the solutions are such that the 
numerical and the approximate solution agree at a position $r_0$, with 
$ 0 < r_0 < r_h$.  
As is to be expected, the approximation improves as $\go$ increases. 

\iffigs
\begin{figure}[bt]
\begin{minipage}[t]{6.5cm}
  \epsfxsize=6.5cm\epsfbox{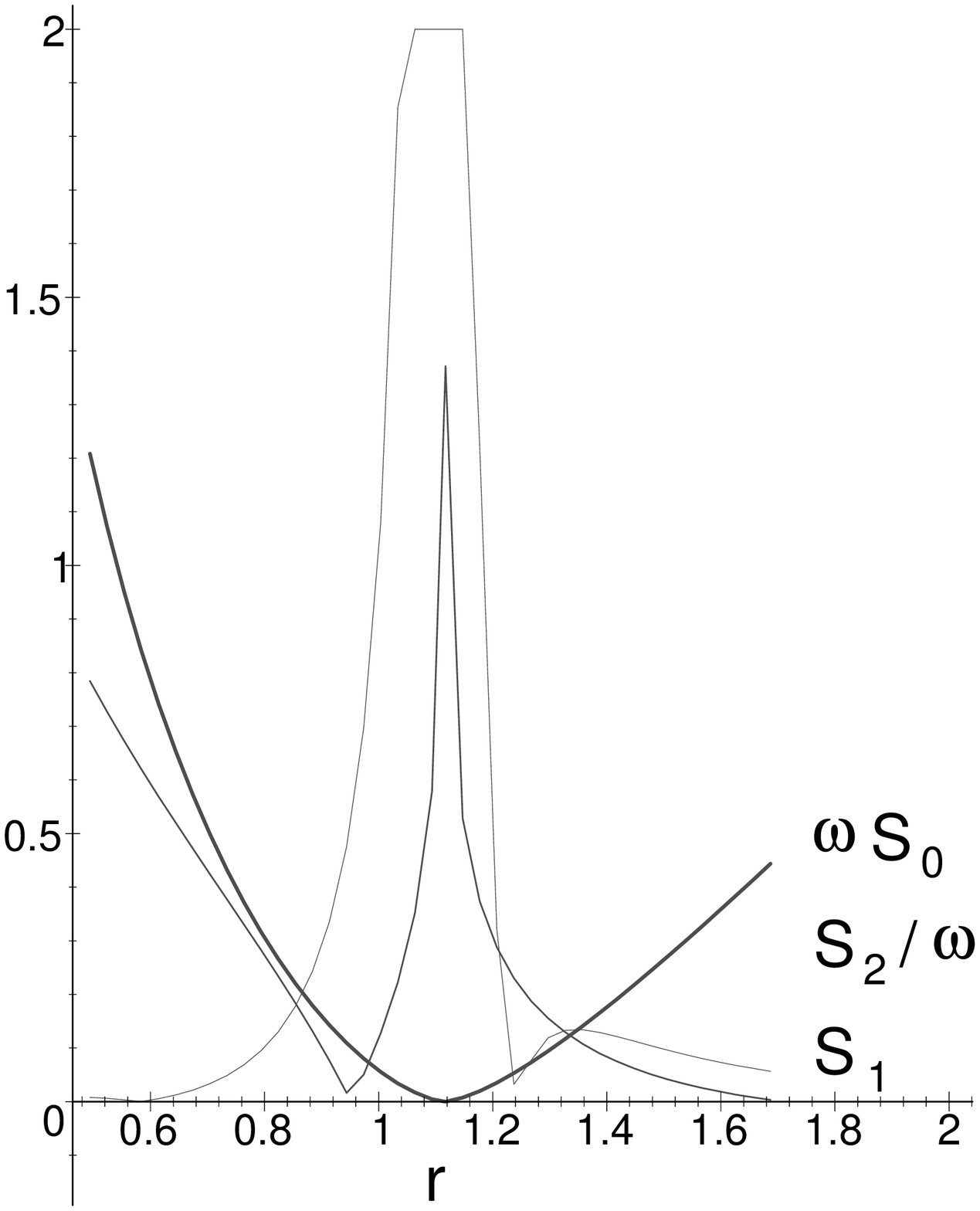}
  \small{{\bf figure 7a:} Comparison of $\go S_0, S_1$ and $S_2/\go$
     for $\go = 1.0$}
\end{minipage}
\hfill
\begin{minipage}[t]{6.5cm}
  \epsfxsize=6.5cm\epsfbox{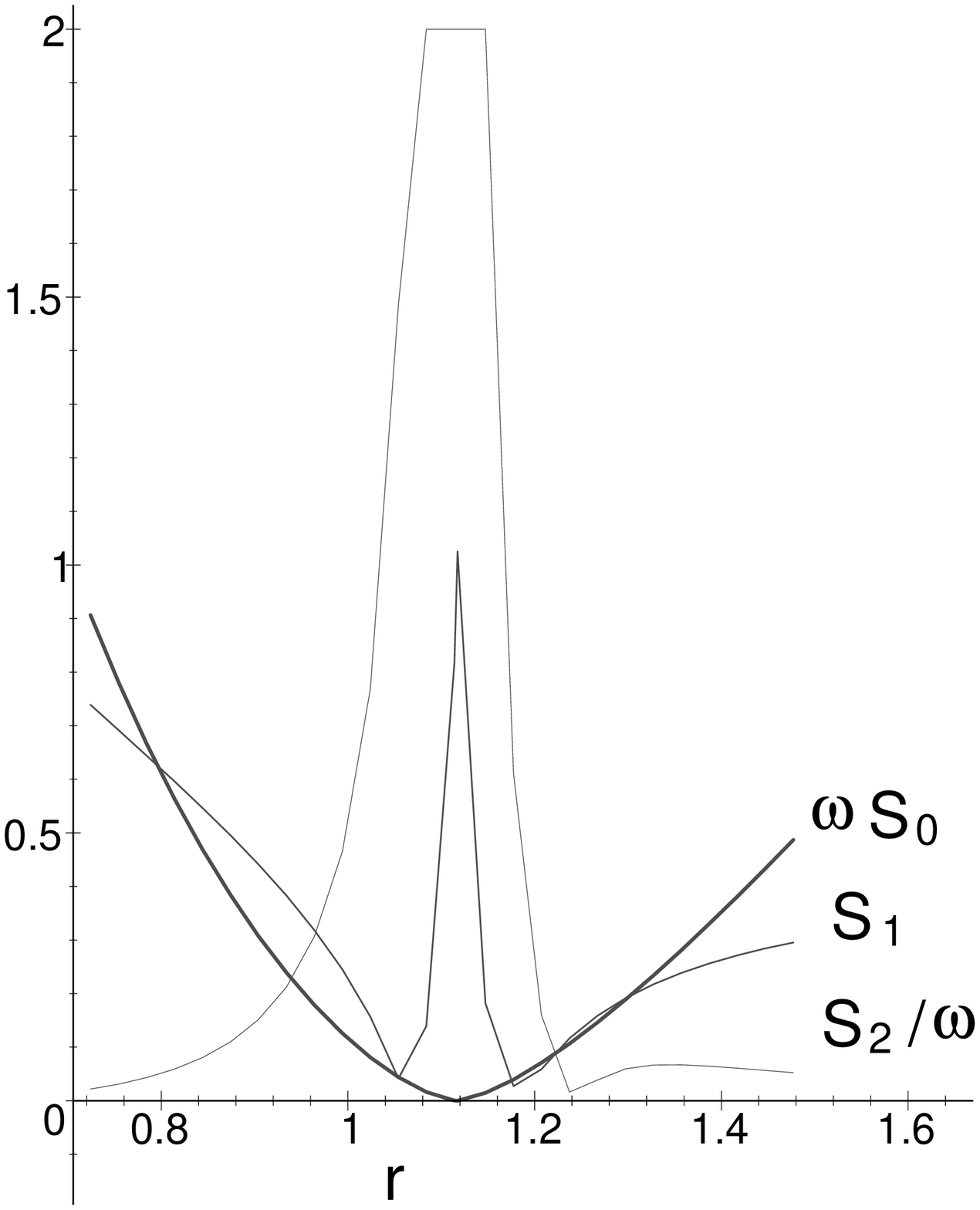}
  \small{{\bf figure 7b:} Comparison of $\go S_0, S_1$ and $S_2/\go$
     for $\go = 1.0$}
\end{minipage}
\end{figure}
\fi

We investigated the range of validity for the approximate solutions 
numerically. 
For the physical optics approximation to be valid the conditions 
(\ref{wkberr}) have to hold up to the term $S_2$. 
The numerical results are illustrated in {\bf figure 7a} 
and {\bf figure 7b} for $\go = 1.0$ and $\go = 2.0$, respectively.  
As a rough estimate for the matching intervals we find the following results: 
If $\go = 1.0$, matching of $\psi_I(r)$ and $\psi_{II}(r)$ and 
$\psi_{II}(r)$ and $\psi_{III}(r)$ seems to be  possible in the intervals 
$[r_h - \mbox{tol},\: 0.8]$ and $[1.4,\: r_h + \mbox{tol}]$. 
Here tol is calculated so that the argument of the Airy function remains 
small, i.e. $\mbox{tol} = \sqrt[3]{-\frac{\epsilon^2}{a}}$.
For $\go = 2.0$ the corresponding intervals are 
$[r_h - \mbox{tol},\: 0.75]$ and $[1.4,\: r_h + \mbox{tol}]$.

At $r = \infty$ the WKB approximation predicts a phase shift of 
$\gd = \frac{\pi}{2}$ between the ingoing and the outgoing mode. 
For large $r$ the ``WKB phase shift'' $\gd_{WKB}(r)$ is calculated for 
the global WKB approximation and extracted from the data for the 
numerical solution.   
The phase shift at $r = 20.0$ for the $\go = 1.0$ data is 
$\gd_{WKB} = 89.54^\circ$. The numerically obtained phase shift 
$\gd_{num}$ is $\gd_{num} = 68.32^\circ$. 
For $\go = 2.0$ the WKB phase shift is equal to $\gd_{WKB} = 89.82^\circ$ 
and the numerical value is $\gd_{num} = 78.71^\circ$.   
Furthermore the WKB approximation suggests that 
the amplitudes of the ingoing and the outgoing modes decay as 
$\sqrt{|C_o|^2 + |C_i|^2}/(2 \: r)$. 
The modulus of the amplitudes of the ingoing and the outgoing mode are 
equal $|C_o| \: =\: |C_i|$, that is the reflection coefficient is equal 
to one. 

For the special repulson data $\ga_3 = \ga_4 = 0$ we can gain some insight 
into the scattering behaviour even for regions of the parameters where the 
physical optics approximation breaks down (i.e. for small $\go$ and 
$l \neq 0$). 

To investigate the asymptotic behaviour of the wave function (\ref{wa1}) 
we use the expansion of the Kummer function for large $r$ 
(see \cite{AbrSte64}). 
Truncation of the asymptotic series after the first term yields 
\bea 
\label{anasym} 
 \psi(r) &\sim& (2 \go r)^s  
 \left[ \me^{-i \go r} \: 
     \frac{(-2i \go r)^{n-s-1}}{\gC(n+s+1)} 
     +
        \me^{i \go r} \:
     \frac{(2i\go r)^{-n-s-1}}{\gC(-n+s+1)} 
           \right] \nonumber \\
      &\sim& \frac{1}{r} \: \sin 
      \left( 
        \go r \:+\: 2 M \go \: \ln 2 \go r \:-\: \frac{\pi}{2} s 
        \:+\: \mbox{arg} \left[ \gC( 1+ s - 2 i M \go) \right] 
      \right).  \nonumber  
\eea
Therefore the modulus of the amplitudes of the ingoing and the outgoing mode 
are still equal. 
But we find a phase shift of 
$\gd = - \frac{\pi}{2} s 
  \:+\: \mbox{arg} \left[ \gC( 1+ s - 2 i M \go) \right].$  
In particular the phase shift is no longer independent of the physical 
data of the repulson. 
In order to get a feeling where the physical optics approximation ``leaks'' 
we apply it to the repulson with $\ga_3 = \ga_4 = 0$. The potential is in 
this case given by 
\be
\label{q0}
  Q_0(r) \:=\: -1 - \frac{4 M}{r} + \frac{s (s+1) }{r^2}.
\ee
Substitution of an asymptotic expansion of $S_0^+(r) \: \sim \: r + 2M \ln r$ 
into (\ref{wkb123}c) yields 
\be
\label{wkbasym}
  \psi_{III}(r) \:=\: \frac{1}{r} \: \sin 
      \left( 
        \go r \:+\: 2 M \go \: \ln r \:+\: \frac{\pi}{4}  
      \right).   
\ee
The physical optics approximation predicts total 
reflection and a phase shift of $\frac{\pi}{2}$ independent of the 
repulson data. The phase shift is constant, i.e. it contains the zeroth 
order contribution only. 
Let us expand the phase shift of the analytical solution in powers of $\go$. 
\bea
  \gd &=& - \frac{\pi}{2} s \:+\: \mbox{arg} \: \gC(1 + s -2iM \go) 
      \: \sim \: \frac{\pi}{4} \:+\: \frac{\pi}{2} \go \:+\: \mbox{arg} \: 
      \gC(1 + s -2iM \go) \nonumber \\
      &\sim& \frac{\pi}{4} \:+\: \go \: (2 M + 1) + \go^2 \: | 1 - 2Mi| \: 
      ( - \sin 4M - 2 M \cos M), \nonumber 
\eea
where we approximated $s$ by $ s \sim - \frac{1}{2} + \go$. 
That is the WKB approximation predicts the zeroth order term in the 
phase shift. 
However, the exact solution also contains first and second order 
contributions. 
This shows that although the physical optics approximation represents the 
functional behaviour nicely it does not approximate the numerical value 
of the phase shift to a high precision. 
This agrees with the observations we made in the case $\ga_3$ and $\ga_4$ 
unequal to zero.  

To summarize: we studied the behaviour of a scalar test--particle in the 
metric background of a repulson. It has a curvature singularity at a 
finite distance $r_h$ and a coordinate singularity at $r = 0$. 
Although the metric changes at $r_h$ 
from a Lorentzian one to one with a Euclidean signature there is nothing 
peculiar about the position $r_h$ of the curvature singularity at the 
quantum level. In contrary, near the coordinate singularity at the origin 
the particle feels a potential barrier and gets reflected. In terms of 
a semi--classical approximation, i.e. high frequency of the ingoing 
particle, there is an equal incoming and outgoing flux, i.e. 
the scattering matrix is unitary. The 
particle is phase shifted by $\frac{\pi}{2}$. 
The numerical data indicate -- as expected -- that the WKB approximation 
does not suit very well for incoming particles with low frequency. 
For special values of the repulson data ($\ga_3 = \ga_4 = 0$) we succeeded 
in working out the asymptotic behaviour analytically. We again find total 
reflection but the phase shift for a low frequency incoming particle 
depends on the physical data of the repulson. 
In particular the scattering behaviour of the scalar test--particle 
at the repulson supports the conjectures that gravitational singularities 
might be smeared out quantum mechanically. 

There are other indications that credit should be given to these naked 
singularities. First of all -- as they are repulsive in nature -- 
they are not a complete desaster from the cosmic censorship point of 
view. 
The singularity is not shielded by a horizon but instead by an effective 
repulsive barrier at which scalar test--particles bounce off. 
In addition they very often correspond to solutions of a higher dimensional 
Kaluza--Klein like theory which usually have different properties. 
Singularities are resolved or naked singularities correspond to black holes  
in higher dimensions. It might be possible to establish 
some relations between these solutions and their properties. 
Unfortunately -- although the repulsons can be associated to solutions 
of 5 dimensional low energy effective string theory -- they do not 
correspond to black holes. At least not, if we take the definition of 
a black hole seriously. They correspond to extremal dilatonic configurations 
which represent space--times with timelike singularities  
\cite{HorMar95, HolWil92, GibWil86}. 

We think it would be very interesting 
to study non--extremal repulson configurations to which one can associate 
black holes and in the sequel entropy, temperature and so far. 
Even more interesting would be to investigate the behaviour of a wave 
packet travelling towards the repulsive barrier. The wave packet would 
serve as a model of a test--string which consists of an infinite number 
of modes. But for that it is necessary to treat the partial differential 
equation (\ref{pde}) in an appropriate manner, which we leave for further 
investigations.


\section*{Acknowledgment}\addcontentsline{toc}
{section}{Acknowledgment}

One of the authors (H. R. H) is indebted to W. Glei{\ss}ner 
and S. Theisen for valuable discussions and for reading parts  
of the manuskript. She also likes to thank G. Gibbons for some 
enlightening remarks on black holes.

The work has been supported in part by the UK Particle Physics 
and Astronomy Research Council and by the Deutsche 
Forschungsgemeinschaft (DFG). 


\end{document}